\documentclass[12pt,a4paper,superscriptaddress]{revtex4}
\usepackage{graphicx}
\usepackage{graphics}
\usepackage{amsmath}
\usepackage{dcolumn}
\usepackage{amssymb}
\usepackage{bm}
\usepackage[latin1]{inputenc}
\newcommand{\be}{\begin{equation}}
\newcommand{\ee}{\end{equation}}
\newcommand{\bea}{\begin{eqnarray}}
\newcommand{\eea}{\end{eqnarray}}

\newcommand{\pa}{\partial}

\begin{document}

\title{On the gauge dependence of the effective potential for Horava-Lifshitz-like theories}

\author{A. M. Lima, J. R. Nascimento, A. Yu. Petrov, R. F. Ribeiro}

\affiliation{Departamento de F\'{\i}sica, Universidade Federal da 
Para\'{\i}ba\\
Caixa Postal 5008, 58051-970, Jo\~ao Pessoa, Para\'{\i}ba, Brazil}
\email{jroberto,petrov,rfreire@fisica.ufpb.br}

\begin{abstract}
We calculate the one-loop effective potential for 
Horava-Lifshitz-like QED with an arbitrary critical exponent within different approaches and discuss its gauge dependence.
\end{abstract}
\maketitle

The effective potential is known to be a key object of the quantum field theory allowing to make conclusions about many issues related to the low-energy effective behaviour of the corresponding theory such as spontaneous symmetry breaking, phase transitions, Green functions and many other aspects. It has been studied within numerous contexts and for different field theory models. Certainly, it is interesting to study the effective potential also in the theories where Lorentz symmetry is broken, in particular, in the theories with time-space asymmetry, that is, those ones characterized by different orders in space and time derivatives. First attempts of studies of such theories \cite{Anselmi}, motivated further by the Horava gravity concept \cite{Horava}, called attention to investigation of different properties of other Horava-Lifshitz-like (HL-like) field theory models, with two main lines of their study are, first, their renormalization aspects \cite{renorm}, second, their effective potential which has been studied for different models including HL-like QED, HL-like Yukawa model and different scalar theories in \cite{EPold,ourHor,ourHor1,ourHor2}. At the same time, one could note that, within studies of the effective potential in the HL-like QED \cite{ourHor,ourHor1,ourHor2}, a special gauge has been employed, that is, the HL-like generalization of the Feynman gauge. Certainly, it simplifies the calculations essentially. However, the problem of the gauge dependence of the effective potential is still open. In this paper we try to answer this problem.
 
Our starting point is the Lagrangian of the scalar QED  with an arbitrary $z$ is \cite{ourHor}
\bea
\label{lasqed}
L&=&\frac{1}{2}F_{0i}F_{0i}+(-1)^z\frac{1}{4}F_{ij}\Delta^{z-1} F_{ij}+D_0\phi
(D_0\phi)^*-
D_{i_1}D_{i_2}\ldots D_{i_z}\phi(D_{i_1}D_{i_2}\ldots D_{i_z}\phi)^*.
\eea
where $D_0=\pa_0-ieA_0$, $D_i=\pa_i-ieA_i$ is a gauge covariant
derivative. For the sake of the simplicity, we suggest that there is no self-coupling of the matter field, the theory is massless, and the critical exponents for scalar and gauge fields are the same (the generalization for the case of their difference is straightforward, as well as for the case of the massive theory).
Here we used slightly different definitions in comparing with the papers \cite{ourHor,ourHor1,ourHor2}, for the convenience. Our signature is $(-+++)$.

We introduce the canonical momenta conjugated to $A_i,\phi,\phi^*$ respectively:
\bea
\Pi_i=F_{0i};\quad\, \pi=(\pa_0+ieA_0)\phi^*=(D_0\phi)^*, \quad\, \pi^*=(\pa_0-ieA_0)\phi=D_0\phi.
\eea
At this time we note the presence of the primary constraint
\bea
\Phi^{(1)}=\Pi_0\simeq 0.
\eea
By didactic reasons, we introduce an intermediate object $L[\Pi,\pi]$, that is, the Lagrangian where the velocities are expressed in terms of momenta:
\bea
L[\Pi,\pi]&=&\frac{1}{2}\Pi_i\Pi_i+(-1)^z\frac{1}{4}F_{ij}\Delta^{z-1} F_{ij}+\pi\pi^* -
D_{i_1}D_{i_2}\ldots D_{i_z}\phi(D_{i_1}D_{i_2}\ldots D_{i_z}\phi)^*.
\eea
The Hamiltonian density is defined as
\bea
H=\Pi_i\dot{A}_i+\pi\dot{\phi}+\pi^*\dot{\phi}^*-L[\Pi,\pi].
\eea
Its explicit form, after one integration by parts, is
\bea
H&=&\frac{1}{2}\Pi_i\Pi_i+\pi\pi^*-A_0(\pa_i\Pi_i-ie(\pi\phi-\pi^*\phi^*))-(-1)^z\frac{1}{4}F_{ij}\Delta^{z-1} F_{ij} +\nonumber\\&+&
D_{i_1}D_{i_2}\ldots D_{i_z}\phi(D_{i_1}D_{i_2}\ldots D_{i_z}\phi)^*.
\eea
The secondary constraint has the role of the Gauss law:
\bea
\label{seccon}
\Phi^{(2)}=\{\Phi^{(1)},H\}=\pa_i\Pi_i+\rho,
\eea
where $\rho=-ie(\pi\phi-\pi^*\phi^*)$ is a charge density. One can conclude that there is no constraints more (indeed, the time dependence in our theory is just the same as in the usual QED). So, our Hamiltonian density is rewritten as
\bea
H&=&\frac{1}{2}\Pi_i\Pi_i+\pi\pi^*-A_0\Phi^{(2)}-(-1)^z\frac{1}{4}F_{ij}\Delta^{z-1} F_{ij} +\nonumber\\&+&
D_{i_1}D_{i_2}\ldots D_{i_z}\phi(D_{i_1}D_{i_2}\ldots D_{i_z}\phi)^*.
\eea
Now, let us follow the algorithm of \cite{BN}. We introduce the transverse projector $P^{\perp}_{ij}$ and the longitudinal one $P^{\parallel}_{ij}$ (we note that for a pure HL-like QED without scalar matter, the Hamiltonian analysis has been performed in \cite{Gaona}):
\bea
P^{\perp}_{ij}=\delta_{ij}-\frac{\pa_i\pa_j}{\Delta}; \quad\, P^{\parallel}_{ij}=\frac{\pa_i\pa_j}{\Delta}.
\eea 
It is clear that $P^{\perp}_{ij}P^{\perp}_{jk}=P^{\perp}_{ik}$, $P^{\perp}_{ij}P^{\parallel}_{jk}=0$, and $\pa_iP^{\perp}_{ij}=0$. Other important properties of the projectors are also valid.
Then, we introduce the transverse and longitudinal momenta:
\bea
\Pi_i^t=P^{\perp}_{ij}\Pi_j,\quad\, \Pi_i^l=P^{\parallel}_{ij}\Pi_j.
\eea
In the same manner we can introduce transverse and longitudinal fields $A^t_i$ and $A^l_i$. We note that the transverse field $A^t_i$ is invariant under the gauge transformations.
It is easy to show that $F_{ij}\Delta^{z-1} F_{ij}=-2A^t_i\Delta^z A^t_i$ (up to the additive total derivative). So, we arrive at the following form for the Hamiltonian density:
\bea
\label{ham}
H&=&\frac{1}{2}\Pi_i^{t}\Pi_i^{t}+\frac{1}{2}\Pi_i^l\Pi_i^l+\pi\pi^*-A_0\Phi^{(2)}+\frac{1}{2}A_i^t(-\Delta)^z A_i^t +\nonumber\\&+&
D_{i_1}D_{i_2}\ldots D_{i_z}\phi(D_{i_1}D_{i_2}\ldots D_{i_z}\phi)^*.
\eea
We note that the secondary constraint now can be rewritten as 
$\Phi^{(2)}=-(\pa_i\Pi_i^l+\rho)$, so, it involves only longitudinal part of the vector field as it must be.
Now, let us carry out the same trick as in \cite{Boyan}, that is, we make the change of variables
\bea
\phi\to\tilde{\phi}=e^{ieR}\phi, \quad\, \phi^*\to\tilde{\phi}^*=\phi^*e^{-ieR}.
\eea
Here $R$ does not depend on $\phi,\phi^*$. It is clear that the conjugated momenta are defined as
\bea
\tilde{\pi}=e^{-ieR}\pi, \quad\, \tilde{\pi}^*=e^{ieR}\pi^*,
\eea
thus, $\pi\pi^*=\tilde{\pi}\tilde{\pi}^*$. Then, it is clear that
$$
(\pa_i+ieA_i)\phi=(\pa_i+ieA_i)(e^{-ieR}\tilde{\phi})=(\pa_i\tilde{\phi}+ie[A_i-\pa_i R]\tilde{\phi})e^{-ieR}.
$$
If one changes $R(x)=-\int d^d y(\pa_j A^l_j(y))G(x-y)$, with $\nabla^2G(x-y)=-\delta^d(x-y)$, and $A^l_j$ is a longitudinal part of $A_j$ (so, $\frac{\pa_i\pa_j}{\nabla^2}A^l_j=A^l_i$), one has $A_i^l-\pa_i R=0$, therefore, 
\bea
(\pa_i+ieA_i)\phi=[(\pa_i+ieA_i^t)\tilde{\phi}]e^{-ieR}.
\eea
Using the mathematical induction method together with the relation $A_i^l-\pa_i R=0$, one can show that for any integer $n$,
\bea
(\pa_{i_n}+ieA_{i_n})\ldots(\pa_{i_1}+ieA_{i_1})\phi=[(\pa_{i_n}+ieA^t_{i_n})\ldots(\pa_{i_1}+ieA^t_{i_1})\tilde{\phi}]e^{-ieR}.
\eea
Therefore, our Hamiltonian density is
\bea
\label{ham1}
H&=&\frac{1}{2}\Pi_i^{t}\Pi_i^{t}+\frac{1}{2}\Pi_i^l\Pi_i^l+\tilde{\pi}\tilde{\pi}^*+A_0\Phi^{(2)}+\frac{1}{2}A_i^t(-\Delta)^z A_i^t +\nonumber\\&+&
[(\pa_{i_n}+ieA^t_{i_n})\ldots(\pa_{i_1}+ieA^t_{i_1})\tilde{\phi}][(\pa_{i_n}+ieA^t_{i_n})\ldots(\pa_{i_1}+ieA^t_{i_1})
\tilde{\phi}]^*.
\eea

Now, it is the time to remember that our aim consists in calculation of the one-loop effective potential. So, we make the shift $\tilde{\phi}\to\tilde{\Phi}+\tilde{\phi}$, $\tilde{\phi}^*\to\tilde{\Phi}^*+\tilde{\phi}^*$, and suggest that, first the field $A_i$ is a purely quantum one, second, the background fields $\tilde{\Phi},\tilde{\Phi}^*$ are constants (which in terms of the original fields is equivalent to the condition that $\Phi\Phi^*=\tilde{\Phi}\tilde{\Phi}^*$ is a constant), so, $\pa_i\tilde{\Phi}=\pa_i\tilde{\Phi}^*=0$. Restricting ourselves by the terms of the second order in quantum fields and integrating by parts where it is necessary, we find the following Hamiltonian density:
\bea
H&=&\frac{1}{2}\Pi_i^t\Pi_i^t+\frac{1}{2}\Pi_i^l\Pi_i^l+\tilde{\pi}\tilde{\pi}^*+\frac{1}{2}A_i^t(-\Delta)^z A_i^t +\tilde{\phi}(-\Delta)^z\tilde{\phi}^*+\nonumber\\&+&
e^2 \Phi\Phi^* A^t_i (-\Delta)^{z-1}A^t_i-A_0\Phi^{(2)}.
\eea
Now, the transverse and longitudinal parts are completely separated.
The condition $\Phi^{(2)}\simeq 0$ emerges as a consequence of the corresponding constraint.
Then, we proceed as in \cite{BN}: we can solve the secondary constraint (\ref{seccon}) as
\bea
\Pi_i^l(x)=\pa_i\int d^3y G(x-y)\rho(y),
\eea
where $G(x-y)$ is a Green function for the Laplace operator, such as $\nabla^2G(x-y)=-\delta(x-y)$, and $\rho (y)$ is a (gauge invariant) charge density. In this case, we can eliminate the longitudinal momenta $\Pi^l_i$  so that
\bea
\int d^3 x\frac{1}{2}\Pi_i^l\Pi_i^l&=&\frac{1}{2}\int d^3xd^3y \rho(x)G(x-y)\rho(y).
\eea
So, our Hamiltonian, on the surface of the constraint, takes the form
\bea
H&=&\int d^3x\Big(\frac{1}{2}\Pi_i^t\Pi_i^t+\frac{1}{2}A_i^t(-\Delta)^z A_i^t +
e^2 \Phi\Phi^* A^t_i (-\Delta)^{z-1}A^t_i+\nonumber\\&+&
\tilde{\pi}\tilde{\pi}^*+\tilde{\phi}(-\Delta)^z\tilde{\phi}^*\Big)+ 
\frac{1}{2}\int d^3xd^3y \rho(x)G(x-y)\rho(y).
\eea
It is clear that the dynamics of scalar fields is completely factorized out, and this Hamiltonian yields the well-known contribution to the one-loop effective potential \cite{ourHor,ourHor1,ourHor2}:
\bea
\label{ep10}
U^{(1)}_t=d\int\frac{d^dk}{(2\pi)^d}[\vec{k}^{2z}+2e^2\Phi\Phi^*\vec{k}^{2z-2}]^{1/2},
\eea
whose result has been found in \cite{ourHor} to be
\bea
\label{ep11}
U^{(1)}_t&=&-\frac{d\pi^{\frac{d-1}{2}}}{{4(2\pi)^d}}(2e^2\Phi\Phi^*)^{\frac{d+z}{2}}\frac{\Gamma\Big(-\frac{d+z}{2}\Big)\Gamma\Big(\frac{d+z-1}{2}\Big)}{\Gamma\Big(\frac{d}{2}\Big)}.
\eea
So, we reproduced the result found in \cite{ourHor} for a HL-like analogue of the Feynman gauge. In other words, it is clear within this formalism that the coupling of the gauge field to quantum scalar fields contributes only to the gauge dependent part.

Finally, we conclude that the only contribution to the effective potential is just (\ref{ep10}). Actually, we have shown that this result does not depend on the gauge choice.

Within other approach, we start again with the expression (\ref{lasqed}) and note that it can be in principle rewritten in terms of the real fields $\phi_1$ and $\phi_2$ such as
\bea
\phi=\frac{\phi_1+i\phi_2}{\sqrt{2}}, \quad\, \phi^*=\frac{\phi_1-i\phi_2}{\sqrt{2}}.
\eea
However, we postpone introduction of $\phi_1$, $\phi_2$ up to the certain step since the formulation with $\phi,\phi^*$ is much more convenient for the quantum calculations. Moreover, actually the only place where we will use the fields $\phi_1$ and $\phi_2$ rather than $\phi,\phi^*$ now will be the gauge condition.

Then, we introduce the following analogue for the $R_{\xi}$ gauge (cf. \cite{BN}) by modifying gauge-fixing Lagrangian from the form used in \cite{BN} to
\bea
L_{gf}=\frac{1}{2\xi}(-1)^z[(-1)^z\Delta^{-\frac{z-1}{2}}\pa_0A_0+\Delta^{\frac{z-1}{2}}\pa_iA_i+e\epsilon_{ab}v_a\phi_b]^2.
\eea
The $v_i$ actually is an isovector in a two-dimensional space. We choose $v_i=(v,0)$, so, $\epsilon_{ab}v_a\phi_b=v\phi_2=v\frac{\phi-\phi^*}{i\sqrt{2}}$. Also, we choose the background $\Phi_i=(\Phi,0)$ to provide $\epsilon_{ij}v_i\Phi_j=0$ (this relation is required by the gauge invariance reasons, cf. \cite{BN}) which in terms of the fields $\phi,\phi^*$ will mean that the background scalar field is real, $\Phi^*=\Phi$.
Since the gauge transformations are as usual
\bea
\delta A_{0,i}=\pa_{i,0}\omega,\quad\, \delta\phi_a=-e\epsilon_{ab}\omega\phi_b,
\eea
We should also introduce a Lagrangian for the corresponding ghosts $c,c^{\prime}$:
\bea
L_{gh}=\frac{1}{\sqrt{\xi}}c[(-1)^z\Delta^{-\frac{z-1}{2}}\pa_0\pa_0+\Delta^{\frac{z+1}{2}}+e^2v\Phi ]c^{\prime}.
\eea
The total Lagrangian of the gauge field will take the form (cf. \cite{ourHor,ourHor1,ourHor2}):
\bea
L_{gauge}&=&\frac{1}{2}A_0\big(-\Delta+\frac{1}{\xi}\pa^2_0(-\Delta)^{-(z-1)}\big)A_0-\pa_0A_0\pa_iA_i(1-\frac{1}{\xi})-\nonumber\\&-&
\frac{1}{2}A_j[\pa^2_0+(-\Delta)^z]A_j+
\frac{1}{2}(1-\frac{1}{\xi})\pa_iA_i(-\Delta)^{z-1}\pa_jA_j+\nonumber\\&+&
\frac{1}{2\xi}e^2v^2\phi^2_2(-1)^z+\frac{1}{\xi}ev[\Delta^{-\frac{z-1}{2}}\pa_0A_0+(-1)^z\Delta^{\frac{z-1}{2}}\pa_iA_i]\phi_2,
\eea
where $\phi_2=\frac{\phi-\phi^*}{2i}$.
Since our aim consists in calculating the effective potential, we as usual carry out the background-quantum splitting by the rule $\phi\to\Phi+\phi$, $\phi^*\to\Phi^*+\phi^*$ (with $\Phi,\Phi^*$ are the background fields, and $\phi,\phi^*$ are quantum ones). We get the following quadratic Lagrangian of quantum fields from the scalar sector:
\bea
L_{sc}&=&\pa_0\phi\pa_0\phi^*+ie[(\pa_0A_0)-(-\Delta)^{z-1}\pa_iA_i](\Phi\phi^*-\Phi^*\phi)+e^2A_0A_0\Phi\Phi^*-\phi(-\Delta)^z\phi^*-\nonumber\\&-&
e^2A_i(-\Delta)^{z-1}A_i\Phi\Phi^*.
\eea
After we impose the condition of reality for the background, $\Phi^*=\Phi$, and introduce the $\phi_1,\phi_2$ fields as above, we get
\bea
L_{sc}&=&\frac{1}{2}[\pa_0\phi_1\pa_0\phi_1-\phi_1(-\Delta)^z\phi_1+\pa_0\phi_2\pa_0\phi_2-\phi_2(-\Delta)^z\phi_2+2e^2\Phi^2A_0A_0-\nonumber\\&-&2e^2\Phi^2A_i(-\Delta)^{z-1}A_i]+(\sqrt{2})e\Phi [(\pa_0A_0)-
(-\Delta)^{z-1}(\pa_iA_i)]\phi_2.
\eea
We sum $L_{sc}$, $L_{gauge}$ and $L_{gh}$. As a result, the total quadratic action is
\bea
L_{total}&=&-\frac{1}{2}\phi_1[\pa^2_0+(-\Delta)^z]\phi_1-\frac{1}{2}\phi_2[\pa^2_0+(-\Delta)^z+\frac{e^2v^2}{\xi}(-1)^{z-1}]\phi_2+\nonumber\\&+&\frac{1}{2}A_0\Big(\frac{\xi^{-1}\pa^2_0+(-\Delta)^z+2e^2\Phi^2(-\Delta)^{z-1}}{(-\Delta)^{z-1}}\Big)A_0- \nonumber\\&-&
\frac{1}{2}A_j[\pa^2_0+(-\Delta)^z+2e^2\Phi^2(-\Delta)^{z-1}]A_j+\nonumber\\&+&
((\sqrt{2})e\Phi +\frac{1}{\xi}ev\Delta^{{-\frac{z-1}{2}}})[(\pa_0A_0)-
(-\Delta)^{z-1}(\pa_iA_i)]\cdot\phi_2+\nonumber\\
&+&
\frac{1}{2}(1-\frac{1}{\xi})\pa_iA_i(-\Delta)^{z-1}\pa_jA_j-\pa_0A_0\pa_iA_i(1-\frac{1}{\xi})\nonumber\\&+&
c[(-1)^z\Delta^{-\frac{z-1}{2}}\pa_0\pa_0+\Delta^{\frac{z+1}{2}}+e^2v\Phi]c^{\prime}.
\eea
Here we reabsorbed the factor $\frac{1}{\sqrt{\xi}}$ into redefinition of the ghosts. We note that the ghost contribution is completely factorized, as it must be in the one-loop order, and we will consider it in the final step.

In principle, we can write down the non-ghost contribution to the corresponding one-loop effective potential as a trace of the logarithm of some operator:
\bea
\Gamma^{(1)}_{\phi,A}=\frac{i}{2}{\rm tr}\ln\left(
\begin{array}{cccc}
-\Box_z & 0 & 0 & 0\\
0 & -\Box_z+\frac{1}{\xi}e^2v^2(-1)^z & T_0 & T_i\\
0 & -T_0 & Q & \pa_0\pa_i(1-\frac{1}{\xi})\\
0 & -T_i & \pa_0\pa_i(1-\frac{1}{\xi}) & -H_{ij} 
\end{array}
\right).
\eea
Here $\pa^2_0+(-\Delta)^z\equiv \Box_z$, $P_z=\Box_z+2e^2\Phi^2(-\Delta)^{z-1}$, $Q=-\Delta+\frac{1}{\xi}\pa^2_0(-\Delta)^{-(z-1)}+2e^2\Phi^2$, $H_{ij}=\delta_{ij}P_z+(1-\frac{1}{\xi})(-\Delta)^{z-1}\pa_i\pa_j$, 
$T_0=((\sqrt{2})e\Phi +\frac{1}{\xi}ev\Delta^{{-\frac{z-1}{2}}})\pa_0$, $T_i=-((\sqrt{2})e\Phi +\frac{1}{\xi}ev\Delta^{{-\frac{z-1}{2}}})(-\Delta)^{z-1}\pa_i$. Already at this step, the field $\phi_1$ (corresponding to the first line/column) completely decouples, so, one rests with
\bea
\Gamma^{(1)}_{\phi,A}=\frac{i}{2}{\rm tr}\ln\left(
\begin{array}{ccc}
-\Box_z+\frac{1}{\xi}e^2v^2(-1)^z & T_0 & T_i\\
-T_0 & Q & \pa_0\pa_i(1-\frac{1}{\xi})\\
-T_i & \pa_0\pa_i(1-\frac{1}{\xi}) & -H_{ij} 
\end{array}
\right).
\eea
This is a result for the one-loop effective potential in an arbitrary gauge. 
In principle, one can reduce even this determinant through the following formula for the determinant of the block matrix:
\bea
\ln\det\left(\begin{array}{cc}
A & B\\
C & D
\end{array}
\right)=\ln \det A+\ln\det (D-CA^{-1}B),
\eea 
with, if we choose $A=-\Box_z+\frac{1}{\xi}e^2v^2(-1)^z$, the first term of this logarithm yields a mere constant and thus can be thrown away. However, the second term, in the case of the arbitrary gauge, is very complicated (the same situation takes place in \cite{BN}).

So, let us choose some gauge in which our one-loop effective potential is radically simplified. It is easy to see that, first,
the cancellation of the "mixed" scalar-vector term requires an essentially nonlocal condition 
\bea
((\sqrt{2})e\Phi +\frac{1}{\xi}ev\Delta^{{-\frac{z-1}{2}}})[(\pa_0A_0)-
(-\Delta)^{z-1}(\pa_iA_i)]=0.
\eea
Actually, this is a generalization of the Feynman gauge which can be treated as an equation on the $A_0,A_i$ (really, it is weaker than the usual Feynman-like gauge condition $(\pa_0A_0)-
(-\Delta)^{z-1}(\pa_iA_i)=0$). In what rests, we can put also $\xi=1$. We rest with the quadratic action of quantum fields (except of ghosts):
\bea
L_{total}&=&\frac{1}{2}[\pa_0\phi_1\pa_0\phi_1-\phi_1(-\Delta)^z\phi_1+\pa_0\phi_2\pa_0\phi_2-\phi_2(-\Delta)^z\phi_2+e^2v^2\phi^2_2(-1)^z]+\nonumber\\
&+&\frac{1}{2}A_0\big(-\Delta+\pa^2_0(-\Delta)^{-(z-1)}+2e^2\Phi^2\big)A_0-\nonumber\\&-&
\frac{1}{2}A_j[\pa^2_0+(-\Delta)^z+2e^2\Phi^2(-\Delta)^{z-1}]A_j.
\eea
It is clear that the contribution to the one-loop effective action from the scalar fields is trivial since it does not involve any background fields (note that $v$ is a constant, not a field), and from the gauge fields one has
\bea
\Gamma^{(1)}_{\phi,A}=\frac{i}{2}(d+1){\rm tr}\ln[\pa^2_0+(-\Delta)^z+2e^2\Phi^2(-\Delta)^{z-1}].
\eea
It is clear that at $z=1$, the expression is Lorentz invariant, and the usual result for the QED is restored.
After the Fourier transform and Wick rotation,  we have
\bea
\label{ep1}
\Gamma^{(1)}_{\phi,A}=\frac{1}{2}(d+1)\int\frac{dk_{0E}d^dk}{(2\pi)^{d+1}}\ln(k^2_{0E}+\vec{k}^{2z}+2e^2\Phi^2\vec{k}^{2z-2}),
\eea
which, with use of \cite{ourHor} is
\bea
\label{h1}
\Gamma^{(1)}_{\phi,A}&=&-\frac{(d+1)\pi^{\frac{d-1}{2}}}{{4(2\pi)^d}}(2e^2\Phi^2)^{\frac{d+z}{2}}\frac{\Gamma\Big(-\frac{d+z}{2}\Big)\Gamma\Big(\frac{d+z-1}{2}\Big)}{\Gamma\Big(\frac{d}{2}\Big)}.
\eea
The only difference is the overall factor $d+1$ instead of $d$ in \cite{ourHor,ourHor1}. However, this is a natural impact of difference of the gauge choice.

Now, recalling \cite{ourHor1}, we can briefly describe the dependence of this result on $d$ and $z$. It is easy to see that when $d+z=2n+1$ is odd, the one-loop effective potential is essentially finite. Moreover, if in this case the $n$ is even, the factor $-\Gamma\Big(-\frac{d+z}{2}\Big)=-\Gamma(-n-\frac{1}{2})$ in (\ref{h1}) is positive, therefore the effective potential is non-negative having the minimum at $\Phi=0$, and if $n$ is odd, the effective potential is negative and the theory is unstable at one loop. At the same time, if $d+z=2l$ is even, the one-loop effective potential diverges and requires an introduction of a corresponding counterterm, that is, the self-coupling of the scalar field, with additional one-loop contributions \cite{ourHor2}.

It remains to treat the ghost contribution to the one-loop effective action. In this case, it is nontrivial being equal to
\bea
\Gamma^{(1)}_{gh}=-\frac{i}{2}{\rm tr}\ln[(-1)^z\Delta^{-\frac{z-1}{2}}\pa_0\pa_0+\Delta^{\frac{z+1}{2}}+e^2v\Phi], 
\eea
or, as is the same,
\bea
\Gamma^{(1)}_{gh}=-\frac{i}{2}{\rm tr}\ln[\pa_0\pa_0+(-1)^z\Delta^z+e^2v\Phi(-1)^z\Delta^{\frac{z-1}{2}}], 
\eea
which after Fourier transform and Wick rotation yields
\bea
\Gamma^{(1)}_{gh}=-\frac{1}{2}\int\frac{d^d\vec{k}dk_0}{(2\pi)^{d+1}}\ln[k^2_0+\vec{k}^{2z}+e^2v\Phi(-1)^{\frac{3z-1}{2}}
|\vec{k}|^{z-1}]. 
\eea
To avoid the problems with reality of the expression, we can suggest that $(-1)^{\frac{3z-1}{2}}=\pm 1$ (that is, $z$ must be odd), with the sign of $v$ is chosen in an appropriate manner.
It remains to integrate which we can do following the lines of \cite{ourHor}. Afterwards, we arrive at
\bea
\label{h2}
\Gamma^{(1)}_{gh}=\frac{1}{(4\pi)^{\frac{2d+z+1}{4}}}\frac{1}{z+1}\frac{\Gamma(\frac{2d+z-1}{2z+2})}{\Gamma(\frac{2d+z-1}{4})}
\Gamma(-\frac{1}{2}-\frac{2d+z-1}{2z+2})(e^2|v\Phi|)^{\frac{d+z}{z+1}}.
\eea
This contribution diverges if $\frac{2d+z-1}{z+1}=2n-1$, with $n$ is a non-negative integer (in particular, if $z=1$, it corresponds to an odd $d$ as it must be). The whole result is a sum of (\ref{h1}) and (\ref{h2}). We note that, first, it diverges at certain values of $d$ and $z$, second, while (\ref{h1}) does not depend on $v$, the (\ref{h2}) essentially depends on it, which means that these two contributions have essentially distinct structure.

Let us compare the results obtained within two approaches, that is, (\ref{ep11}) and the sum of (\ref{h1}) with (\ref{h2}), respectively. It is clear that the last result is obtained in a some special gauge, that is, the analogue of $R_{\xi}$ gauge allowing for removal of non-diagonal terms of the action by paying a price of introducing the extra parameters $v$ and $\xi$, and, further, the nontrivial coupling the ghosts to the scalar field. It is clear that this gauge is much more generic than the usual Feynman-like gauge used in the first part of the paper. However, if we suggest that we impose several special restrictions on this parametes, that is, choose $\xi=1$ to remove the non-diagonal terms in the purely gauge sector, and $v=0$ to remove the ghost-matter coupling  together with its consequence, that is, the conbtribution (\ref{h2})-- remind that within usual gauges, which do not involve scalar fields, this coupling does not arise -- the results, that is, (\ref{ep11}) and (\ref{h1}), will have exactly the same functional form. The only difference is in the overall factor ($d$ or $d+1$), which is caused by the fact that while within the first manner of calculation we absorbed $A_0$ (so-called "scalar photon") into the charge density $\rho$ which has no contribution at the one-loop order, within the second manner we treated $A_0$ on the same base as the physical $A_i$ components which yields a contribution similar to that one of $A_i$. Therefore, we conclude that, if we restrict ourselves to purely physical variables, that is, throw away the contribution of a non-physical $A_0$, the results will coincide. Actually, the difference of results within two methods is caused by the fact that, in the $R_{\xi}$ gauge, the result for the one-loop effective potential is strongly gauge dependent. Nevertheless, the physical variables should be gauge independent.

We considered two different approaches to study of the one-loop effective potential in the HL-like QED. Within the first of them, we implemented the gauge invariant physical variables and obtained the one-loop potential expressed in these variables. The result coincided with the previous result of \cite{ourHor}. Within the second one, we used a special gauge, that is, $R_{\xi}$ gauge known as an efficient tool in simplification of the classical action. We showed that using of this gauge with an appropriate fixation of the free parameters allows to obtain a result which differs from that one of \cite{ourHor} only by a numerical factor, plus some extra contribution generated by ghosts which do not decouple in this case because of the unusual structure of the gauge-fixing function. 

{\bf Acknowledgements.} This work was partially supported by Conselho
Nacional de Desenvolvimento Cient\'{\i}fico e Tecnol\'{o}gico (CNPq).
A. Yu. P. has been supported by the CNPq project 303438-2012/6.

\end{document}